%% file: document.tex
\documentclass{sig-alternate}
\pdfoutput=1
\usepackage{balance}

\usepackage{tikz}
\usetikzlibrary{calc,arrows,positioning,decorations.pathreplacing}

\usepackage[linesnumbered,vlined,ruled]{algorithm2e}
\usepackage{microtype}
\usepackage{graphicx}
\usepackage{color}
\usepackage{caption}
\usepackage{subcaption}
\usepackage{xspace}
\usepackage{pgfplots}
\usepackage{pgfplotstable}
\usepackage{booktabs}
\usepackage{times}

\usepackage[binary-units=true]{siunitx}

\usepackage{braket}

\usepackage{microtype}

\newtheorem{mydef}{Definition}

\usepackage{fmtcount}

\usepackage{tabularx}
\usepackage{multirow}

\usepackage{wrapfig}

\usepackage{makecell}

\usepackage{afterpage}
\newlength{\oldtextfloatsep}

\makeatletter
\renewcommand{\paragraph}{%
  \@startsection{paragraph}{4}%
  {\z@}{0.5ex \@plus 1.0ex \@minus .2ex}{-0.5em}%
  {\normalfont\normalsize\bfseries}%
}
\makeatother

\newcommand{\abbrev}[1]{\textsc{\MakeLowercase{#1}}\xspace}
\newcommand{\GRAPHITE}{\abbrev{GRAPHITE}}
\newcommand{\CPU}{\abbrev{CPU}}
\newcommand{\RDBMS}{\abbrev{RDBMS}}

\newcommand{\LS}{\abbrev{LS}}
\newcommand{\LSFULL}{\LS-traversal }
\newcommand{\FI}{\abbrev{FI}}
\newcommand{\FIFULL}{\FI-traversal }
\newcommand{\TGI}{\abbrev{TGI}}

\newcommand{\SIMD}{\abbrev{SIMD}}

\newcommand{\SSD}{\abbrev{SSD}}

\newcommand{\GMSs}{\textsc{\MakeLowercase{GMS}}'s }

\newcommand{\SQL}{\abbrev{SQL}}
\newcommand{\SECRET}{\abbrev{SAP HANA}}
\newcommand{\GRAPHCHI}{GraphChi }

\newcommand{\TURBOGRAPH}{TurboGraph}
\newcommand{\GBASE}{GBase }

\newcommand{\GMS}{\abbrev{GMS}}
\newcommand{\RDF}{\abbrev{RDF}}
\newcommand{\XML}{\abbrev{XML}}

\newcommand{\EP}{\abbrev{EP}}
\newcommand{\CR}{\abbrev{CR}}

\newcommand{\PA}{\abbrev{PA}}
\newcommand{\LJ}{\abbrev{LJ}}

\newcommand{\SK}{\abbrev{SK}}
\newcommand{\TW}{\abbrev{TW}}
\newcommand{\OR}{\abbrev{OR}}

\begin{document}

\title{GRAPHITE: An Extensible Graph Traversal Framework \\ for Relational Database Management Systems}

\numberofauthors{3}
\author{
   \begin{tabular*}{0.9\textwidth}{@{\extracolsep{110pt}}cc}
           Marcus Paradies,
           Wolfgang Lehner
           &Christof Bornh{\"o}vd
    \end{tabular*}\\[10pt]
    \begin{tabular*}{0.8\textwidth}{@{\extracolsep{90pt}}cc}
           \affaddr{Database Technology Group}&\affaddr{SAP Labs, LLC}\\
           \affaddr{Technische Universit\"at Dresden}&\affaddr{Palo Alto, CA 94304, USA}\\
           \email{\normalsize{\texttt{\textbf{marcus.paradies@gmail.com}}}}&\email{\normalsize{\texttt{\textbf{christof.bornhoevd@sap.com}}}}\\
           \email{\normalsize{\texttt{\textbf{wolfgang.lehner@tu-dresden.de}}}}
    \end{tabular*}
}

\maketitle
\begin{abstract}
Graph traversals are a basic but fundamental ingredient for a variety of graph algorithms and graph-oriented queries. To achieve the best possible query
performance, they need to be implemented at the core of a database management system that aims at storing, manipulating, and querying graph data. Increasingly,
modern business applications demand native graph query and processing capabilities for enterprise-critical operations on data stored in relational database
management systems. In this paper we propose an extensible graph traversal framework (\GRAPHITE) as a central graph processing component on a common storage
engine inside a relational database management system.

We study the influence of the graph topology on the execution time of graph traversals and derive two traversal algorithm implementations specialized for
different graph topologies and traversal queries. We conduct extensive experiments on \GRAPHITE for a large variety of real-world graph data sets and input
configurations. Our experiments show that the proposed traversal algorithms differ by up to two orders of magnitude for different input configurations and
therefore demonstrate the need for a versatile framework to efficiently process graph traversals on a wide range of different graph topologies and types of
queries. Finally, we highlight that the query performance of our traversal implementations is competitive with those of two native graph database management
systems.
\end{abstract}

\input{1-introduction}
\input{2-framework}
\input{3-kernels}
\input{4-ls_traversal}
\input{5-fi_traversal}
\input{6-topology_optimizations}
\input{7-evaluation}
\input{8-related}
\vspace{20pt}
\input{9-conclusion}


\balance

\bibliographystyle{abbrv}
\bibliography{bibliography}
\end{document}

%% file: 1-introduction.tex
\section{Introduction}

\begin{figure}[t!]
\begin{subfigure}{0.45\linewidth}
    \includegraphics[width=0.9\textwidth,natwidth=610,natheight=642]{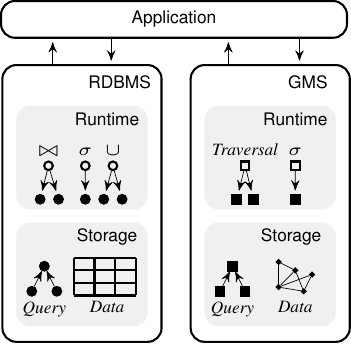}
    \caption{Separated \RDBMS and \GMS with specialized storage engines and processing via application.}
    \label{subfig:first}
\end{subfigure}
\hspace{\fill}
\begin{subfigure}{0.4\linewidth}
    \includegraphics[width=0.9\textwidth,natwidth=610,natheight=642]{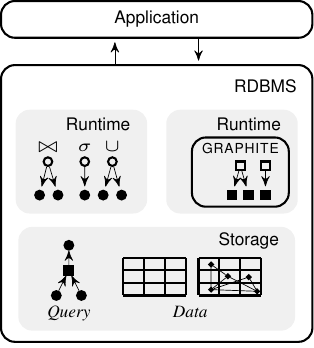}
    \caption{\RDBMS with integrated graph processing capabilities and common storage engine.}
    \label{subfig:second}
\end{subfigure}
\vspace{-5pt}
\caption{Architecture alternatives for graph processing.}
\label{fig:architecture_overview}
\vspace{-15pt}
\end{figure}

Evermore, enterprises from various domains, such as the financial, insurance, and pharmaceutical industry, explore and analyze the connections between data
records in traditional customer-relationship management and enterprise-resource-planning systems. Typically, these industries rely on mature \RDBMS technology
to retain a single source of truth and access. Although graph structure is already latent in the relational schema and inherently represented in foreign key
relationships, managing native graph data is moving into focus as it allows rapid application development in the absence of an upfront defined database schema.
Specifically, novel and traditional business applications leverage the advantages of a graph data model, such as schema flexibility and an explicit
representation of relationships between data records. Although these business applications mainly operate on graph-structured data, they still require direct
access to the relational base data. \newline \indent Existing solutions performing graph operations on business-crit\-ic\-al data either use a combination of
\SQL and application logic or employ a graph management system (\GMS) such as Neo4j~\cite{Neo} or Sparksee~\cite{Martinez-Bazan2007}, or distributed graph systems, such as
GraphLab~\cite{Low2012} or Apache Giraph~\cite{Gir}. For the first approach, relying only on \SQL typically results in poor execution performance caused by the
functional mismatch between a traversal algebra~\cite{Rodriguez2011} and the relational algebra. Even worse, the relational query optimizer is not \emph{graph-aware}
i.e., it does not keep statistics about the graph topology nor about graph query patterns, and therefore is likely to construct a suboptimal execution plan. The
other alternative is to process the data in a native \GMS to hurdle the unsuitability of the relational algebra to express complex graph queries in an \RDBMS.

Figure~\ref{subfig:first} depicts a traditional system landscape with an \RDBMS and a \GMS located next to each other and orchestrated at application level. A
\GMS is superior to an \RDBMS for complex graph processing as it provides a natural understanding of a graph data model, a rich set of graph processing
functionality, and optimized data structures for fast data access. Especially scenarios that do not require accessing the most recent data snapshot nor combine
operations from different data models into \emph{cross-data-model operations} can be handled by \GMSs efficiently. Cross-data-model operations combining data
from various data models, i.e., relational, text, spatial, temporal, and graph however will play a key role for graph analytics in the future~\cite{Abadi2014}. For
example, a clinical information system stores data from patient records in an \RDBMS. Graph analytics on a knowledge graph of patient records and their
relationships to each other help physicians to improve diagnostics and identify complex co-morbidity conditions. Such a medical knowledge graph contains not
only information about the relationships between diagnoses and patients, but also text data from patient records and temporal information about prescriptions.

In this paper we propose the seamless integration of graph processing functionality into an \RDBMS sharing a common storage engine as depicted in
Figure~\ref{subfig:second}. Located next to a relational runtime stack in the same system, a graph runtime with a set of graph operators provides native support
for querying graph data on top of a common relational storage engine. For the context of this paper, we focus on graph traversals as they are a vital component
of every \GMS and the foundation for a large variety of graph algorithms, such as finding shortest paths, detecting connected components, and answering
reachability queries. \newline \indent We introduce \GRAPHITE, a traversal framework that provides an extensible set of logical graph traversal operators and
their corresponding implementations. Similar to the distinction between a logical and a physical layer in a relational runtime, \GRAPHITE also provides a set of
logical operators and a set of corresponding physical implementations. In the context of this paper we propose two traversal implementations optimized for
in-memory columnar \RDBMS but argue that the general concept of a traversal framework can be extended with specialized traversal implementations and cost models
for row-oriented or even disk-based \RDBMS. \GRAPHITE operates on a physical column group model (cf. Figure~\ref{fig:mapping}). We summarize our main
contributions as follows:
\vspace{-5pt}
\begin{itemize}
  \setlength{\itemsep}{0pt}
  \item We introduce \GRAPHITE as a modular and extensible foundation of a traversal framework inside an \RDBMS, which allows seamlessly reusing existing
  physical data structures and deploying of novel traversal implementations.
  \item We present two different implementations of the traversal operator, a naive level-synchronous (\LS), and a novel fragmented-incremental (\FI) traversal
  algorithm that is superior to the naive approach for specific graph topologies and traversal queries.
  \item We conduct an extensive experimental evaluation for a large variety of real-world data sets and traversal queries, and show an execution time
  improvement of our \FIFULL by up to two orders of magnitude compared to the \LSFULL for certain graph topologies and traversal queries. Moreover, we show that
  the query performance of our implementations is competitive with those of two native graph database management systems.
\end{itemize}%
\vspace{-5pt}

\noindent The remainder of this paper is structured as follows: in Section~\ref{sec:framework}, we describe \GRAPHITE as the foundation of the traversal
operator that we present in Section~\ref{sec:kernels}. We detail the two physical implementations of the traversal operator in Sections~\ref{sec:lstraversal}
and \ref{sec:fitraversal}, respectively. A set of topology-aware clustering techniques that can be applied to both physical implementations is presented in
Section~\ref{sec:improvements}. In Section~\ref{sec:evaluation} we provide an extensive experimental evaluation of our traversal implementations. Finally, we
discuss related work in Section~\ref{sec:related} before we conclude the paper in Section~\ref{sec:summary}.

%% file: 2-framework.tex
\section{Graph Traversal Framework}
\label{sec:framework}
\noindent \GRAPHITE is a general and extensible framework that allows easy deploying, testing, and benchmarking of traversal implementations on top of a common
relational storage engine of an \RDBMS. It provides a common interface for traversal configuration parameters and is tightly integrated with a unified,
\emph{graph-aware} controller infrastructure leveraging a comprehensive set of available graph statistics. We show that the query optimizer of an \RDBMS can
benefit from having extensive information about the graph topology to choose the best traversal operator for a given traversal query.

\begin{figure}
\setlength{\textfloatsep}{0pt}
\begin{subfigure}[l]{0.2\textwidth}
\includegraphics[width=1.0\textwidth,natwidth=610,natheight=640]{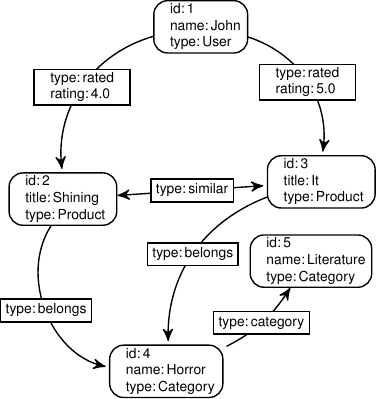}
\vspace{-5pt}
\subcaption{Example graph.}
\end{subfigure}
\hspace{5pt}
\begin{subfigure}[r]{0.22\textwidth}
\input{vertextable}
\subcaption{Vertex column group.}
\vspace{5pt}
\input{edgetable}
\subcaption{Edge column group.}
\end{subfigure}
\vspace{-5pt}
\caption{Mapping of a property graph to column groups.}
\vspace{-10pt}
\label{fig:mapping}%
\end{figure}

\paragraph*{Graph Model and Physical Representation} \GRAPHITE provides as logical data model a property graph model. The property graph data model has emerged
as the de-facto standard for general purpose graph processing in enterprise environments~\cite{Rodriguez2010}. It represents multi-relational directed graphs where
vertices and edges can have assigned an arbitrary number of attributes in a key/value fashion. \newline \indent We store a property graph using a common storage
infrastructure with the relational runtime stack in two column groups, one for the vertices and one for the edges, respectively. A column group represents a
column-oriented data layout, where a new attribute can be added by appending a new column to the column group~\cite{Abadi2007}. Null values in sparsely populated
columns can be compressed through run-length-based compression techniques~\cite{Abadi2006}. Additionally, the evaluation of column predicates allows a seamless
combination of relational predicate filters with the actual traversal operation. Figure~\ref{fig:mapping} depicts an example graph and its mapping to two column
groups. We map each vertex and edge to a single entry in the column group and each attribute to a separate column. Each vertex has a unique identifier as the
only mandatory attribute.

\paragraph*{Traversal Configuration} In the following, we introduce a formal notion of the graph traversal operation, its input parameters, and the expected
output.
\begin{mydef}(\textbf{Traversal Configuration}) Let $G=(V,E)$ be a directed, multi-relational graph, where $V$ refers to the set of vertices and $E \subseteq (V
\times V)$ refers to the set of edges. We define a traversal configuration $\rho$ as a tuple $\rho=(S,\varphi,c,r,d)$ composed of a set of start vertices $S
\subseteq V$, an edge predicate $\varphi$, a collection boundary $c$, a recursion boundary $r$, and a traversal direction $d$. A graph traversal
$\tau_{G}(\rho)$ is a unary operation on $G$ and returns a set of visited vertices $R \subseteq V$.
\label{def:traversal}
\end{mydef}
\noindent We represent each vertex in $S$ by its unique vertex identifier. The edge predicate $\varphi$ defines a propositional formula consisting of atomic
attribute predicates that can be combined with the logical operators $\land$, $\lor$, and $\neg$. For each edge $e \in E$, the traversal algorithm evaluates
$\varphi$ and appends matching edges to the working set of \emph{active edges}. Further, it receives a recursion boundary $r \in \mathbb{N}^+$ that defines the
maximum number of levels to traverse. To support unlimited traversals or transitive closure calculations, the recursion boundary can be infinite ($\infty$). The
collection boundary $c \in \mathbb{N}$ specifies the level of the traversal from where to start collecting discovered vertices. For $c=0$, we add all start
vertices to the result. For any traversal configuration, the condition $c \leq r$ must hold. The traversal direction $d \in \Set{\rightarrow,\leftarrow}$
specifies the direction to traverse the edges. A forward traversal~($\rightarrow$) traverses edges from the source vertex to the target vertex, a backward
traversal~($\leftarrow$) traverses edges in the opposite direction. The traversal operation outputs a set of vertices that have been visited in the boundaries
defined by $c$ and $r$.

\begin{figure}[t]
\centering
\begin{minipage}{0.12\textwidth}
\scriptsize
\includegraphics[width=1.0\textwidth,natwidth=610,natheight=640]{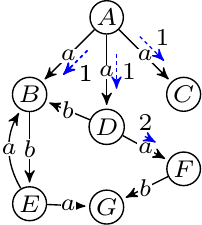}
\end{minipage}
\hspace{-5pt}
\begin{minipage}{0.35\textwidth}
\centering
\pgfplotstabletypeset[string type,
                      columns/config/.style={column name=Traversal Configuration,column type=r},
                      columns/result/.style={column name=Result,column type=l},
                      every head row/.style={before row=\toprule,after row=\midrule},
                      every last row/.style={after row=\bottomrule},
                      every row no 1/.style={},
                      every row no 3/.style={},
                      font=\scriptsize]
{
config result
\,$(\Set{A},\text{'\emph{type=a}'},0,1,\rightarrow)$ $\Set{A,B,C,D}$
\,$(\Set{A},\text{'\emph{type=a}'},0,1,\rightarrow)$ $\Set{A,B,C,D}$
\,$(\Set{A},\text{'\emph{type=a}'},1,1,\rightarrow)$ $\Set{B,C,D}$
\,$(\Set{A},\text{'\emph{type=a}'},2,2,\rightarrow)$ $\Set{F}$
\,$(\Set{A},\text{'\emph{type=a}'},1,\infty,\rightarrow)$ $\Set{B,C,D,F}$
\,$(\Set{E},\text{'\emph{type=b}'},2,2,\leftarrow)$ $\Set{D}$
\,$(\Set{A},\text{'\emph{type=a $\vee$ type=b}'},2,2,\rightarrow)$ $\Set{E,F}$
}
\end{minipage}
\vspace{-5pt}
\caption{Traversal configurations and result sets.}
\label{fig:exampleConfigs}
\end{figure}

\paragraph*{Formal Description} We define a traversal by a totally ordered set $P$ of path steps, where each path step describes the transition between two
traversal iterations. Path steps are evaluated sequentially according to the total ordering in $P$. We determine the number of path steps by the recursion
boundary $r$. Formally, we define a graph traversal operation based on the mathematical notion of sets and their basic operations \emph{union} and
\emph{complement}. Each path step $p_i \in P$ with $1\leq i\leq r$ receives a set of vertices $D_{i-1}$ discovered at level $i-1$ and returns a set of adjacent
vertices $D_{i}$. Initially, we assign the set of start vertices to the set of discovered vertices ($D_0 = S$). In the following, we define the transformation
rules for $p_i$ with $i>0$.
\begin{equation}
D_{i}^{\rightarrow} = \Set{ v | \exists u \in D_{i-1}: e = (u,v) \in E \wedge
    eval(e,\varphi)}
    \label{eq:normal_forward_transition}
\end{equation}
\vspace{-15pt}
\begin{equation}
D_{i}^{\leftarrow} = \Set{ u | \exists v \in D_{i-1}: e = (u,v) \in
E \wedge eval(e,\varphi)}
    \label{eq:normal_backward_transition}
\end{equation}
\noindent Depending on the traversal direction $d$, we select a different transformation rule. Equation~\ref{eq:normal_forward_transition} presents the
definition for forward traversals ($\rightarrow$), and Equation~\ref{eq:normal_backward_transition} for backward traversals ($\leftarrow$), respectively.
In path step $p_i$ we generate the set of vertices $D_i$ by traversing from each vertex in $D_{i-1}$ over all outgoing/incoming edges matching the edge
predicate $\varphi$. Once the traversal operation finished processing the path step, the vertex set $D_i$ contains all vertices reachable within one hop from
the vertices in $D_{i-1}$ via edges for which $\varphi$ holds. Equation~\ref{eq:final} shows the definition of the resulting vertex set $R_{\tau}$ for a
traversal operation $\tau$.
\begin{equation}
R_{\tau} = \Bigg(\underbrace{\bigcup \limits_{i=c}^{r} D_i}_\text{target
vertices}\Bigg) \setminus \Bigg(\underbrace{
\bigcup
\limits_{i=0}^{c-1} D_i }_\text{visited vertices}\Bigg)
\label{eq:final}
\end{equation}
\noindent Conceptually, the collection boundary $c$ and the recursion boundary $r$ divide the discovered vertices into two working sets.
The set of \emph{visited vertices} contains all vertices that have been discovered before the traversal reached the collection boundary $c$. Vertices within the
set of visited vertices are not considered for the final result, but are necessary to complete the traversal operation. We produce the set of visited vertices
by forming the union of all partial vertex sets $\{D_0,D_1,\ldots,D_{c-1}\}$ from path steps $p_1$ to $p_{c-1}$. The set of \emph{target vertices} refers to the
set of vertices that are potentially relevant for the final result set. To produce the set of target vertices, we union all partial vertex sets
$\{D_c,\ldots,D_{r}\}$ from path steps $p_c$ to $p_r$. To retrieve the final result, we compute the complement between the set of target vertices and the set of
visited vertices. We consider only vertices from the set of target vertices that are not in the set of visited vertices for the final result.
Figure~\ref{fig:exampleConfigs} shows a set of example traversal configurations and their query results for the given graph.

\noindent\textbf{Example.} Traversal configuration $(\Set{A},\text{'\emph{type=a}'},2,2,\rightarrow)$ in Figure~\ref{fig:exampleConfigs} traverses starting from
vertex $A$ on edges of type $a$ and visits all vertices within a distance of 2 from vertex $A$. Here, we only collect discovered vertices in the last path step
$p_2$. Dashed arrows with numbers show the traversed edges and the path step they were discovered in. First, path step $p_1$ transforms the vertex set $D_0 =
\Set{A}$ into the vertex set $D_1 = \Set{B,C,D}$. Next, path step $p_2$ transforms vertex set $D_1$ into vertex set $D_2 = \Set{F}$. Finally, the output for
this example graph and traversal configuration is a vertex set containing vertex $F$ only.

%% file: vertextable.tex
\pgfplotstableset{string type, 
                  font=\tiny, 
                  columns/id/.style={ column name=$id$, column type=O },
                  columns/title/.style={ column name=$title$, column type=T },
                  columns/name/.style={ column name=$name$, column type=r },
                  columns/type/.style={ column name=$type$, column type=r },
                  columns/others/.style={ column name=$\ldots$, column type=O},
                  empty cells with={--}}

\newcolumntype{T}{>{\hfill\arraybackslash}p{8mm}}%
\newcolumntype{O}{>{\centering\arraybackslash}p{1mm}}

\pgfplotstabletypeset[every head row/.style={before row=\toprule,
                                             after row=\midrule},
                      every last row/.style={after row=\bottomrule},
                      every row no 1/.style={},
                      every row no 3/.style={}]
{
id type name title others
1 User John ? {$\ldots$}
2 Product ? {Shining} {$\,$}
3 Product ? {It} {$\ldots$}
4 Category Horror ? {$\,$}
5 Category Literature ? {$\ldots$}
}

%% file: edgetable.tex
\pgfplotstableset{string type,
                  font=\tiny,
                  columns/source/.style={ column name=$V_s$, column type=C },
                  columns/target/.style={ column name=$V_t$, column type=C },
                  columns/type/.style={ column name=$type$, column type=F },
                  columns/rating/.style={ column name=$rating$, column type=F},
                  columns/other/.style={ column name=$\ldots$, column type=C},
                  empty cells with={--}
                  }

\newcolumntype{C}{>{\centering\arraybackslash}p{2mm}}%
\newcolumntype{F}{>{\centering\arraybackslash}p{6mm}}%

\pgfplotstabletypeset[every head row/.style={before row=\toprule,
                                             after row=\midrule}, 
                      every last row/.style={after row=\bottomrule}, 
                      every row no 1/.style={}, 
                      every row no 3/.style={}]
{
source target type rating other
2 3 similar ? {$\ldots$}
2 4 belongs ? {$\,$}
3 4 belongs ? {$\ldots$}
1 3 rated 5.0 {$\,$}
1 2 rated 4.0 {$\ldots$}
4 5 category ? {$\,$}
}

%% file: 3-kernels.tex
\section{Traversal Operators}
\label{sec:kernels}

\begin{figure}[t]
\centering
\includegraphics[width=0.45\textwidth,natwidth=600,natheight=640]{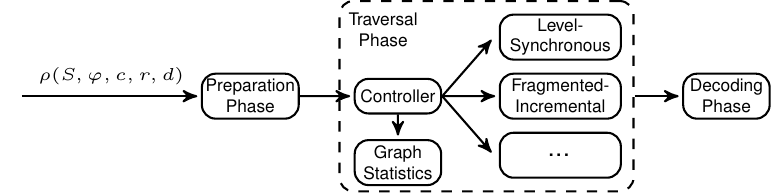}
\vspace{-5pt}
\caption{Processing phases in \GRAPHITE.}
\vspace{-15pt}
\label{fig:workflow}
\end{figure}

\noindent We now discuss the components of \GRAPHITE, as shown in Figure~\ref{fig:workflow}, in detail. It receives a traversal configuration $\rho$ and
processes a traversal in three phases: a \emph{Preparation Phase}, a \emph{Traversal Phase}, and a \emph{Decoding Phase}. All three phases share common
interfaces allowing to easily exchange implementations.

\paragraph*{Preparation Phase} We pass a set of start vertices $S$ to the preparation phase and transform it into a more processing-friendly set-oriented data
structure. If the storage engine leverages dictionary encoding~\cite{Abadi2006}, we consult the value dictionary of the source/target vertex column in the edge
column group, and encode all vertices from $S$ into their internal numerical value code representation. Depending on the traversal direction, we use a different
vertex ID column (either $V_s$ or $V_t$ from Figure~\ref{fig:mapping}). In addition to the actual value encoding, we select active edges that are to be
considered for the traversal operation. Therefore, we push down the edge predicate $\varphi$ to the storage engine to filter out invalid edges. Active edges are
stored in a list that represents the valid and invalid records of a column group. Additionally, transactional visibility is guaranteed by intersecting the
obtained list with visibility information in the current transaction context from the multi-version concurrency control of the \RDBMS. Finally, we pass the list
of active edges to the traversal phase for further processing.

\paragraph*{Traversal Phase} We distinguish two major components in the traversal phase: a set of \emph{traversal operator implementations} and a
\emph{traversal controller}. Within the scope of this paper, we propose two traversal algorithm strategies---\emph{level-synchronous} and
\emph{fragmented-incremental}---and describe them in detail in Sections~\ref{sec:lstraversal} and \ref{sec:fitraversal}, respectively.
Initially, we pass a collection boundary $c$, a recursion boundary $r$, a traversal direction $d$, a set of active edges $E_\textit{a}$, and the encoded vertex
set $S$ to the traversal phase. We select the best traversal operator implementation based on collected graph statistics and the characteristics of the
traversal query. After the traversal operator has finished execution, it returns discovered vertices in a set-oriented data structure.

\paragraph*{Decoding Phase} To translate the internal code representations of discovered vertices back into actual ID values, we consult the value dictionary of
the source/target vertex column for each value code and add the actual value to the final output set. If the storage engine does not leverage dictionary
encoding, the decoding phase can be omitted and the result set directly returned.

\subsection{Strategies and Variations}

\noindent The core component of an implementation of the graph traversal operator is the traversal algorithm. In general, traversal algorithms occur in
variations favoring different graph topologies and traversal queries. While dense graphs with a large vertex outdegree favor a more robust (with respect to
skewed outdegree distribution) level-synchronous traversal algorithm, a very sparse graph with a low average vertex outdegree benefits from a more fine-granular
traversal implementation. We divide the algorithm engineering space for traversal implementations into two dimensions: \emph{traversal strategy} and
\emph{physical reorganization}.

Within the scope of this paper, we propose two traversal algorithms named \LSFULL and \FIFULL in the traversal strategy dimension. The second dimension
describes the physical organization of the edges. We distinguish between a \emph{clustered} physical organization, where edges having the same source vertex are
clustered together in the column, and an \emph{unclustered} physical organization, where edges do not have a particular physical ordering. Both dimensions are
freely combinable with each other. The \LSFULL operates level-synchro\-nously and thereby emits only those vertices within a single traversal iteration that are
adjacent to the vertices from the working set. For each traversal iteration, it reads the complete graph to retrieve adjacent vertices. For sparse graphs, each
edge is accessed possibly multiple times although each edge is only traversed exactly once. In multi-core environments with a large number of available hardware
threads, this overhead can be hidden through parallelized read operations on the graph. For an underutilized database management system with a low query
workload, such a read-intensive implementation can hide the additional cost for reading data multiple times. However, a single traversal query cannot leverage
the full parallelization capabilities in a fully utilized database management system with a high query workload and possibly hundreds of traversal queries
running in parallel. In such a scenario, the \CPU is fully occupied and the overhead for reading the complete graph multiple times for a single traversal query
cannot be concealed anymore. To keep the execution time of a single query low, either more hardware resources have to be added or the resource consumption of a
single query has to be reduced.

The \FIFULL uses less \CPU resources than the \LSFULL algorithm and aims at minimizing the total number of read operations on the complete graph. It processes a
graph in \emph{column fragments} and thereby materializes adjacent vertices immediately after the processing of the respective column fragment.
Column fragments divide a column into logical partitions, where partition sizes can vary within a column. This allows limiting the operation area to those parts
of the graph that are relevant for the given query. We give a detailed description of the \FIFULL in Section~\ref{sec:fitraversal}.

%% file: 4-ls_traversal.tex
\section{LS-Traversal}
\label{sec:lstraversal}
\begin{figure}[t]
\centering
\includegraphics[width=0.4\textwidth,natwidth=610,natheight=640]{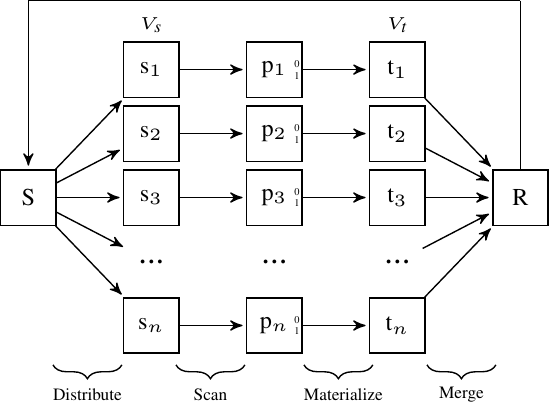}
\vspace{-10pt}
\caption{\LS-traversal algorithm.}
\label{fig:level_synchronous}
\end{figure}

\noindent Figure~\ref{fig:level_synchronous} sketches the execution flow of our \LSFULL implementation. It operates on two columns $V_\textit{s}$ and
$V_\textit{t}$ that represent source and target vertices of edges, respectively. To fully exploit thread-level parallelization, we split columns $V_\textit{s}$
and $V_\textit{t}$ into $n$ equally sized logical partitions of edges. In the following, $s_1,\ldots,s_n$ describe partitions of $V_s$, partitions
$t_1,\ldots,t_n$ correspond to column $V_t$.
An \LSFULL visits vertices in a strict breadth-first ordering and thereby discovers vertices always on the shortest path. Conceptually, we divide an \LSFULL
algorithm into four major algorithmic steps: \emph{Distribute}, \emph{Scan}, \emph{Materialize}, and \emph{Merge}. The distribute step propagates a search
request with the working set of vertices to all $n$ partitions in parallel. Next, each scan worker thread searches for vertices from the working set in its
local partition $s_i$ with $1\leq i \leq n$ and writes search hits into a local position list $p_i$. \newline \indent Each materialization worker thread
receives a local position list $p_i$ and fetches adjacent vertices from the target vertex column $V_\textit{t}$. Subsequently, the merge step collects and
combines all locally discovered adjacent vertices into the vertex set $R$. Finally, the traversal algorithm either terminates and forwards its output to the
decoding phase, or continues with the next traversal iteration.

We sketch our \LSFULL implementation in Algorithm~\ref{list:lstraversal}. Initially, we pass a traversal configuration $\kappa=(S_m,E_\textit{a},c,r,d)$ to the
\LS-traversal. The preparation phase emits a vertex set $S_m$, evaluates the edge predicate $\varphi$ and returns a set of active edges $E_\textit{a}$.
The output of an \LSFULL execution is a set of discovered vertices $R$. We collect intermediate results, such as vertex sets and position lists, either in
space-efficient bit sets or in dense set data structures, depending on the estimated output cardinality of the traversal iteration.

\input{lstraversal}

First, the \LSFULL algorithm analyzes whether the traversal configuration describes a forward or a backward traversal and updates the handles to the columns
accordingly (Line~\ref{alg:line3}). If the collection boundary $c$ is zero, all vertices in $S_m$ are added to the final result $R$ (Line~\ref{alg:line5}).
Initially, we assign the vertex set $S_m$ to the working vertex set $D_w$. The major part of the \LSFULL algorithm describes a single traversal iteration and is
executed at most $r$ times (Lines~\ref{alg:line8}--\ref{alg:line18}). At the beginning of each traversal iteration, we check the working set $D_w$ for
emptiness. If it is empty, no more vertices can be discovered and the execution of the \LSFULL is terminated. During each traversal iteration, we scan the
source vertex column $V_s$ in parallel, and emit matching edges into a position list $P$. During the scan operation, we use the set of active edges
$E_\textit{a}$ to check the validity of the matching edges and filter out invalid edges. In addition, the scan operation modifies the set of active edges by
invalidating all traversed edges. Next, the \LSFULL algorithm materializes adjacent vertices into the working set $D_w$ using the position list $P$. If the
currently active traversal iteration already passed the collection boundary $c$, it adds the discovered vertices from $D_w$ to the result set $R$. Finally, it
passes the working set $D_w$ to the next traversal iteration. The traversal algorithm terminates if either no more vertices have been discovered during the last
traversal iteration or it has reached the recursion boundary $r$.

\paragraph*{Cost Model}
The execution time of the \LSFULL is dominated by the total number of edges in the graph and the number of processed traversal iterations. It has a worst case
time complexity of $\mathcal{O}(r\cdot|E|)$, where $r$ denotes the recursion boundary and $|E|$ refers to the total number of edges in the graph. For each
traversal iteration, it scans the graph for adjacent vertices from a given vertex set.

We provide a query and graph topology-dependent cost model to describe the execution time behavior of the \LS-traversal. The cost of an \LSFULL can be derived
from the number of edges to read and the number of traversal iterations to perform.

\begin{equation}
\mathcal C_{\LS} = \min\{r,\tilde\delta\} \cdot |E| \cdot \mathcal C_e
\label{eq:cost_ls}
\end{equation}

\noindent We define the cost $\mathcal C_{\LS}$ as the composite product of the minimum of the recursion boundary $r$ and the estimated diameter $\tilde\delta$
of the graph, the number of edges $|E|$, and a constant cost $\mathcal C_e$ to read a single edge from main memory in Equation~\ref{eq:cost_ls}.

%% file: lstraversal.tex
\begin{algorithm}[t]
\small
\SetKwInOut{Input}{Input}
\SetKwInOut{Output}{Output}
\SetKwComment{tcp}{\tiny    // }{}
\SetKwComment{tcpblank}{// }{}
\SetKwFunction{scan}{scan}
\SetKwFunction{materialize}{materialize}
\SetKwFunction{swap}{swap}
\Input{ Traversal configuration $\kappa=(S_m,E_\textit{a},c,r,d)$.}
\Output{ Set of discovered vertices $R$.}
\Begin{
\If{$d$ \textit{is backward}}{
    $\swap(V_\textit{s}$,$V_\textit{t});$ \tcp{\tiny Adjust Column Handles}\label{alg:line3}
}
 
\If{$c = 0$}{
    $R \leftarrow S_m$ \tcp{\tiny Add start vertices to result}
    \label{alg:line5}
}
$p \leftarrow 1;
$ $D_w \leftarrow S_m$
\While {$p \leq r$} { \label{alg:line8}
    \If{$D_w = \emptyset$}{
        \Return; \tcp{\tiny No more vertices to discover}
    }

    $P \leftarrow \emptyset$\;
    $V_\textit{s}.\scan(D_w,E_\textit{a},P);$ \tcp{\tiny Parallel scan for $D_\textit{w}$}
    $D_\textit{w} \leftarrow \emptyset;$ \tcp{\tiny Reset working vertex set}
    $V_\textit{t}.\materialize(P,D_\textit{w});$ \tcp{\tiny Materialize vertices from $P$} \label{alg:line13}
    \If{$p \geq c$}{
        $R \leftarrow R \cup D_\textit{w};$ \tcp{\tiny Add vertices from $D_w$ to result $R$}
    }

    $p \leftarrow p + 1$\; \label{alg:line18}
}
}
    \Return $R$;
\caption{\LSFULL}
\label{list:lstraversal}
\afterpage{\global\setlength{\textfloatsep}{\oldtextfloatsep}}
\end{algorithm}

%% file: 5-fi_traversal.tex
\section{FI-Traversal}
\label{sec:fitraversal}
\noindent An \FIFULL attempts to limit read operations of data records to those that are required for creating the final result. Thereby, it preserves the
ability to fully exploit available thread-level parallelism and differs from an \LSFULL in two fundamental ways. First, an \FIFULL materializes adjacent
vertices of a given set of vertices on column fragment granularity instead of column granularity. Therefore, intermediate results can be accessed immediately
and are available before the next scan operation begins. Second, a scan operation searches for adjacent vertices from several unfinished traversal iterations
and outputs results of multiple traversal iterations. Consequently, an \FIFULL does not operate level-synchronously, but instead traverses the graph
incrementally by processing fragments in sequence.

\begin{figure}[t]
\centering
\includegraphics[width=0.45\textwidth,natwidth=610,natheight=640]{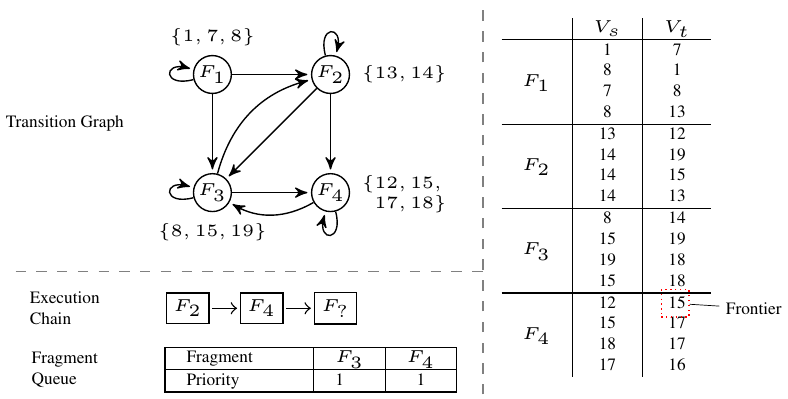}
\captionsetup{skip=5pt}
\caption{Example transition graph and auxiliary data structures.}
\label{fig:transindex}
\end{figure}

We select the next fragment to read with the help of a light-weight, synopsis-based \emph{transition graph index} (\abbrev{TGI}). Conceptually, a \abbrev{TGI}
models a directed graph, where vertices denote fragments and edges represent transitions between them. A transition between two fragments $F_1$ and $F_2$
describes a path of length 2 with an edge $e_1 = (u,v)$ in $F_1$ and an edge $e_2 = (v,w)$ residing in $F_2$. If we read edge $e_1$ in fragment $F_1$ and
proceed with the traversal afterwards, we have to read fragment $F_2$ as well, as it contains edges that extend the traversal path. A fragment has at most one
fragment transition to any other fragment, including to itself. Since not every fragment has a transition to every other fragment in the transition graph, we
only represent directed edges between fragments if there is a transition between them. In addition to fragment transitions, we store light-weight synopses
representing the distinct values of each fragment. A fragment synopsis is stored as a compact bloom filter with bits set for all distinct values present in the
fragment. Figure~\ref{fig:transindex} depicts a transition graph with fragment size 4 for the edge column group on the right side. For example, there is a
fragment transition $F_2 \rightarrow F_4$ with a path $13 \leadsto 12 \leadsto 15$, i.e., $e_1 = (13,12)$ in $F_2$ and $e_2 = (12,15)$ in $F_4$. The fragment
synopses are directly attached to the corresponding fragment, e.g., the fragment synopsis $\{13,14\}$ represents all distinct values in fragment $F_2$.

In addition to the \abbrev{TGI}, we store query-specific runtime information about already processed fragments and fragment candidates in auxiliary data
structures. We keep already processed fragments in a \emph{fragment execution chain} and append to it whenever a new fragment has been selected for execution.
Since we only choose a single fragment at a time, we queue all other generated fragment candidates in a priority-based \emph{fragment queue}. To choose the next
fragment for execution, we select the tail fragment of the execution chain and use the set of newly discovered vertices (the so-called \emph{frontiers}) from
the previous traversal round. Every vertex in the graph can only be a frontier vertex exactly once, i.e., when the vertex is first discovered.

\input{fitraversal}

For the tail fragment, we probe all adjacent fragment synopses with the frontier vertices. If an adjacent fragment matches, i.e. there is a transition between
the tail and the adjacent fragment caused by one or more frontiers, the adjacent fragment is added to the fragment queue. If the fragment is already queued, we
increase its priority. After updating the fragment queue, we remove the fragment with the highest priority and return it to the main traversal algorithm. If
there are no frontier vertices, we immediately remove the fragment with the highest priority. If the fragment queue becomes empty, the traversal terminates.

\textbf{Example.} Let us consider an example traversal starting with vertex 13 as depicted in Figure~\ref{fig:transindex}. We start the traversal at fragment
$F_2$ and emit the newly discovered vertex with id 12. During the fragment selection, we probe all adjacent fragments of $F_2$ (i.e., $F_2$, $F_3$, and $F_4$)
with frontier vertex 12. Since fragment $F_4$ contains vertex 12 in its fragment synopsis, we select it as the next fragment to read. After processing fragment
$F_4$, we emit frontier vertex 15 and select the next fragment to read. Since the probing generates two candidate fragments $F_3$ and $F_4$, we select one of
them and continue the traversal.

After describing the principle workings of the \FIFULL algorithm, we sketch the algorithmic description in Algorithm~\ref{list:fitraversal}. Initially, we pass
a traversal configuration $\kappa$ with a vertex set $S_m$, an edge set $E_{a}$, a collection boundary $c$, a recursion boundary $r$, and a direction $d$ to the
\FIFULL algorithm. It outputs a vertex set $R$ with visited vertices that have been discovered between $c$ and $r$. Since the execution of an \FIFULL is based
on sequential reads of fragments, we parallelize the execution of the scan if necessary and materialize operations within a single column fragment.
An \FIFULL runs in a series of iterations, where we process one fragment per iteration. At the beginning of each iteration, the algorithm
\texttt{getNextFragment} receives a set of frontier vertices and returns the next fragment $F$ to read. A fragment contains the start and end position in the
column and limits the scan to that range. Initially, we pass the set of start vertices as frontiers to \texttt{getNextFragment}. The body of the main loop
performs a scan operation and a materialize operation (Lines~9--10). The scan takes the first $\textit{sFactor}$ working sets from the traversal iterations and
returns matching edges in the corresponding position lists from the vector of position lists $P$. For example, an n-way scan with \emph{sFactor}=2 probes the
column with two vertex sets from two different traversal iterations and returns matching edges into two position lists. Subsequently, newly discovered adjacent
vertices are materialized in a similar multi-way manner as in the scan operation. Depending on the $\textit{mFactor}$, we read out the collected position lists
and add adjacent vertices to the working sets from $D_w$. In addition, newly discovered vertices are added to the set of frontier vertices $Frontiers$. Once the
recursion boundary is reached, the traversal reads and processes all remaining fragments from the fragment queue. If \texttt{getNextFragment} does not return
any more fragments, the traversal terminates and generates the final result according to the given collection and recursion boundaries
(Line~\ref{alg:fi_line13}).

\paragraph*{Candidate Fragment Selection}

\input{transitionGraphIndex}

Algorithm~\ref{list:transition_graph_index} describes in detail how to find the next fragment to read given a set of frontier vertices. It starts with the last
processed fragment and probes adjacent fragments for matching vertices. For each adjacent fragment, we consult its fragment synopsis and compare the frontiers
against it. If a frontier matches, we update the fragment queue accordingly. If the fragment is already in the queue, we increase its priority, otherwise we
insert it. Further, we invalidate vertices in the synopses that triggered the candidate fragment selection. We keep invalidated vertices and their corresponding
fragments in an invalidation list $I$ (Line~6). Finally, we return the fragment with the highest priority from the fragment queue and append it to the execution
chain. Since the fragment synopses are implemented as compact bloom filters, false positive fragments can occur. However, a false positive does not harm the
traversal functionally. There is a tradeoff between space consumption and execution time for the fragment synopses. We evaluate the effect of the size of a
bloom filter in the experimental evaluation. Since the value distribution in fragments might vary, each bloom filter can have a different size depending on the
number of distinct values present in the fragment.

\paragraph*{Cost Model}

The cost model of the \FIFULL is slightly more complex than for the \LSFULL since the calculation of the costs depends on a larger set of input parameters. The
costs of an \FIFULL can be directly related to the number and the size of the accessed fragments. Hence, we can use the chain of read fragments $F_p$ to derive
the cost of the \FI-traversal. The overall cost is the accumulated cost of the reads for all accessed fragments in $F_p$. Consequently, the traversal cost is
not directly dependent on the number of traversal iterations anymore. We define the cost $\mathcal C_\text{FI}$ of an \FIFULL in Equation~\ref{eq:cost_fi} as follows.
\begin{equation}
\mathcal C_{\FI} = \sum_{i=0}^{\min\{r,\tilde\delta\}} (1+p)(\bar{d}_{\text{out}})^i \cdot \xi \cdot \mathcal C_e
\label{eq:cost_fi}
\end{equation}
\noindent The cost depends on the average false positive rate $p$, the average vertex outdegree $\bar{d}_{\text{out}}$, and the fragment size $\xi$. The \FIFULL
is bounded by the minimum of the recursion boundary $r$ and the estimated effective graph diameter $\tilde\delta$. The most important factors effecting the
memory consumption of the \TGI are the size and number of fragments. We can minimize the memory consumption of the \TGI by grouping edges by source vertex (see
edge clustering in Section~\ref{sec:improvements}). Then, each vertex with incoming and outgoing edges contributes exactly once to a single fragment transition.
For equally-sized fragments, we have to choose a fragment size that is as large as the largest vertex out-degree. Therefore, we also propose to provide a
heterogeneous fragment size distribution that is always larger than a predefined minimum fragment size. The upper size is determined automatically by the vertex
out-degree. We discuss these configuration parameters and their performance implications for the \FIFULL in detail in the evaluation in
Section~\ref{sec:evaluation}.

%% file: fitraversal.tex
\setlength{\textfloatsep}{0pt}
\newcommand{\plus}{\raisebox{0.1\height}{\scalebox{.8}{+}}}
\begin{algorithm}[t]
\small
\SetKwInOut{Input}{Input}
\SetKwInOut{Output}{Output}
\SetKwComment{tcp}{    // }{}
\SetKwComment{tcpblank}{// }{}
\SetKwFunction{getNextFragment}{getNextFragment}
\SetKwFunction{nWayScan}{nWayScan}
\SetKwFunction{nWayMaterialize}{nWayMaterialize}
\SetKwFunction{swap}{swap}
\SetKwFunction{generateResult}{generateResult}

\SetKwFunction{extractMin}{extractMin}
\SetKwFunction{hasKey}{hasKey}
\SetKwFunction{increasePrio}{increasePrio}
\SetKwFunction{insert}{insert}

\Input{ Traversal configuration $\kappa=(S_m,E_\textit{a},c,r,d)$.}
\Output{ Set of discovered vertices $R$.}
\Begin{
\If{$d$ \textit{is backward}}{
    $\swap(V_\textit{s}$,$V_\textit{t});$\tcp{\scriptsize Adjust Column
    Handles}\label{alg:line3} }

$D_w[0] \leftarrow S_m$\;
$Frontiers \leftarrow S_m$\;
$sFactor \leftarrow 1$\;
$mFactor \leftarrow 1$\;
\While {$\getNextFragment(Frontiers,F)$}{\label{alg:fi_line8}
    $V_\textit{s}.\nWayScan(F,D_w,E_\textit{a},sFactor,P)$\;\label{alg:fi_line9}
    $V_\textit{t}.\nWayMaterialize(P,mFactor,D_\textit{w},Frontiers)\;$\label{alg:fi_line10}
    \lIf{$sFactor \leq r$}{\plus\plus$sFactor$}
    \lIf{$mFactor < r$}{\plus\plus$mFactor$}
}
$R \leftarrow \generateResult(c,r,D_w)$\; \label{alg:fi_line13}
}
\caption{\FIFULL}
\label{list:fitraversal}
\afterpage{\global\setlength{\textfloatsep}{\oldtextfloatsep}}
\end{algorithm}

%% file: transitionGraphIndex.tex
\begin{algorithm}[t]
\small
\SetKwInOut{Input}{Input}
\SetKwInOut{Output}{Output}
\SetKwComment{tcp}{    // }{}
\SetKwComment{tcpblank}{// }{}
\SetKwFunction{add}{add}
\SetKwFunction{getLast}{getLast}
\SetKwFunction{matches}{matches}

\SetKwFunction{extractMin}{extractMin}
\SetKwFunction{hasKey}{hasKey}
\SetKwFunction{increasePrio}{increasePrio}
\SetKwFunction{insert}{insert}
\SetKwFunction{emp}{empty}

\Input{ Set of frontier vertices $Frontiers$.}
\Output{ Candidate fragment $F$ to read next.}
\Begin{
$F_{last} \leftarrow m\_chain.\getLast()$\;
    \ForEach{outgoing edge $e=(F_{last},F_{cand})$ from $F$}{
        \ForEach{$v \in F$}{
             \If{$F_{cand}.\matches(v) \wedge \neg I.\hasKey(F_{cand},v)$}{\label{alg:tgi_line5}
                $I.\insert(F_{cand},v)$\;\label{alg:tgi_line6}
                \If{$PQ.\hasKey(F_{cand})$} {$PQ.\increasePrio(F_{cand})$}
                \lElse{$PQ.\insert(F)$}\label{alg:fi_line10}
              }
        }
    }
    \If{$\neg PQ.\emp()$}{
        $F \leftarrow PQ.\extractMin()$\;
        $m\_chain.\add(F)$\;
        \KwRet true\;
    }
    \lElse{\KwRet false}
}
\caption{Procedure \texttt{getNextFragment($Frontiers$,$F$)}}
\label{list:transition_graph_index}
\afterpage{\global\setlength{\textfloatsep}{\oldtextfloatsep}}
\end{algorithm}

%% file: 6-topology_optimizations.tex
\section{Topology-Aware Clustering}
\label{sec:improvements}

\noindent The basic implementation of the \LSFULL algorithm does not rely on a particular ordering of the edges in the edge column group. However, to fully
leverage the benefits of a main-memory storage engine, we can use data access patterns that provide a more efficient access to data placed in memory. Therefore,
a physical reorganization of records is a common optimization strategy to reduce data access costs~\cite{Abadi2006}. In the following, we describe two strategies
to further reduce the overall execution time of the \LSFULL algorithm by maximizing the spatial locality of memory accesses to reduce the number of records to
scan.

\paragraph*{Type Clustering}
\setlength{\columnsep}{7pt}%
\begin{wrapfigure}{r}{0.14\textwidth}
  \centering
  \vspace{-15pt}
    \includegraphics[width=0.2\textwidth,natwidth=610,natheight=640]{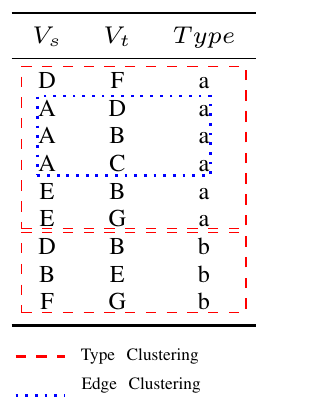}
    \vspace{-15pt}
  \caption{Clustered Edges.}
  \vspace{-10pt}
  \label{fig:clustered_edges}
\end{wrapfigure}
Typically, real-world graph data sets are modeled with a widespread and diverse set of edge types that connect the vertices in the graph. Conceptually, an edge
type describes a subgraph and can be interpreted as a separate layer or view on top of the original data graph. Such multi-relational graphs with multiple edge
types are common in a variety of scenarios, such as product batch traceability, social network applications, or material flows graphs. For example, a product
rating website might store different relationships between entity types \emph{rating}, \emph{user}, and \emph{product}, such as rating relationships, product
hierarchies, and user fellowships. To that end, traversal queries are specific with regard to which parts of the graph they refer to. We propose to arrange
edges sharing the same type physically together, allowing a traversal query to operate directly on the subgraph instead of the entire original graph. Thus, a
graph that comprises $n$ different edge types results in $n$ different subgraphs. A subgraph is associated with an area in the column that contains all edges
forming the subgraph. Figure~\ref{fig:clustered_edges} illustrates an edge column group with two edge types. Here, a traversal query that refers to edges of
type $b$ would only have to scan the corresponding subgraph. The portion of the column for edge type $b$ is indicated by the dashed lower rectangle. If the edge
predicate contains a disjunctive condition, for example to traverse only over edges of type $a$ or $b$, the \LSFULL algorithm automatically splits the scan
operation and unions the partial results thereafter.

\paragraph*{Edge Clustering}

The most fundamental component of a traversal operation is to retrieve the set of adjacent vertices for a given vertex. Therefore, an efficient traversal
implementation must provide efficient access to adjacent vertices located in main memory. To achieve this, we introduce the notion of \emph{topological
locality} in a graph. Topological locality describes a concept for accessing all vertices adjacent to a given vertex $v \in V$. If a neighboring vertex of a
vertex $v$ is accessed, it is likely that all other vertices adjacent of $v$ are accessed also.

We translate topological locality in a graph directly into spatial locality in memory by grouping edges based on their source vertex. Such an edge clustering
increases spatial locality, i.e., all edges sharing the same source vertex are written consecutively in memory. Maximizing spatial locality for memory accesses
results in a better last-level cache utilization and minimizes the amount of data to be loaded from main memory into the last-level cache of the
processor~\cite{Boncz1999}. Figure~\ref{fig:clustered_edges} sketches an example for vertex $A$. All edges having $A$ as source vertex are written consecutively
into the edge column group. To that end, applying first clustering by type and then by edge extends the physical reorganization on a second level.
Especially the materialization step of an \LSFULL algorithm benefits from an increased spatial locality while fetching adjacent vertices from the $V_t$ column
(see Line~\ref{alg:line13} in Algorithm~\ref{list:lstraversal}).

Besides the spatial locality, \emph{column decompression} plays an important role in materializing adjacent vertices. Major in-memory database vendors rely on a
two-level compression strategy. The first level is dictionary encoding, where a value is represented by its numerical value code from the dictionary and stored
in a bit-packed, space-efficient data container. Here, a lightweight, but still notable decompression routine is used to reconstruct the actual value code. If
adjacent vertices are not stored in a consecutive chunk of memory, the decompression routine might decompress unnecessary value codes. A similar behavior can be
observed on the second level of compression, the value-based block compression. Edge clustering allows retrieving blocks of value codes that can be
reconstructed efficiently by leveraging \SIMD instructions.

%% file: 7-evaluation.tex
\section{Evaluation}
\label{sec:evaluation}

\pgfplotsset{
  grid style = {
    dash pattern = on 0.15mm off 1mm,
    line cap = round,
    gray,
    line width = 0.5pt
  }
}

\noindent We evaluate the \LSFULL and the \FIFULL on a diverse set of real-world graphs and for different types of graph queries. In the following, we describe
the environmental setup and provide statistical information about the evaluated data sets. We present an extensive experimental evaluation of the memory
consumption, execution time, cost model, and system-level comparison with two native graph management systems.

\subsection*{Environmental Setup and Data Sets}
\label{sec:setup}
\noindent We have implemented \GRAPHITE as a prototype in the context of the in-memory column-oriented \SECRET database system. Graph data in \SECRET is stored
in two column groups, where each group has its own read-optimized main storage and write-optimized delta storage. Data manipulation operations exclusively
modify the delta storage, which is periodically merged into the main storage. Deletions only invalidate records and affected records are being removed during
the next merge process. Within the scope of this paper, we focus on read-only graphs, but argue that the proposed algorithms could be easily extended to support
dynamic graphs as well (for example by treating the delta storage as a single fragment and by using general visibility data structures, such as a validity
vector, to check for deletions). All values are dictionary-encoded allowing the traversal algorithms to operate on the value codes directly. Initially, we
loaded the data sets into their corresponding vertex and edge column groups, and populate the \TGI. We ran the experiments on a single server machine running
SUSE Linux Enterprise Server 11 (\SI{64}{\bit}) with Intel Xeon X5650 running at \SI{2.67}{\GHz}, 6 cores, 12 threads, \SI{12}{\mega\byte} L3 cache shared, and
\SI{48}{\giga\byte} RAM. For the \LSFULL we leverage full parallelization with 12 threads, for the \FIFULL we use 1 thread to scan a single fragment. To
evaluate our approach on a wide range of different graph topologies, we selected six real-world graph data sets from the domains: social networks~(\OR,\TW,\LJ),
citation networks~(\PA), autonomous system networks~(\SK), and road networks~(\CR). For each data set, we report the number of vertices $|V|$, the number of
edges $|E|$, the average vertex outdegree $\bar{d}_{out}$, the maximum vertex outdegree $\max(d_{out})$, the estimated graph diameter $\tilde\delta$, and the
raw size of the graph in Table~\ref{tab:data}.

{
\pgfplotstableset{ numeric type, reset styles, col sep=comma, row
sep=\\, font=\scriptsize,
columns/datasetid/.style={ string type, column name=\makecell[t]{\emph{ID}} },
columns/vertices/.style={ string type,column name=\emph{$|V|$}, column type=r },
columns/edges/.style={ string type,column name=\emph{$|E|$}, column type=r },
columns/avgoutdegree/.style={ column name=\emph{$\bar{d}_{out}$}, fixed zerofill, precision=1, column type=r },
columns/maxoutdegree/.style={ string type, column name=\makecell[t]{\emph{$\max(d_{out})$}}, column type=r },
columns/diameter/.style={ column name=\makecell[t]{\emph{$\tilde\delta$}}, column type=r, string type},
columns/size/.style={ column name=\emph{Size\,(MB)}, column type=r }, empty
cells with={--}, every head row/.style={before row=\toprule, after row=\midrule},
every last row/.style={after row=\bottomrule} }

\pgfkeys{/pgf/number format/.cd,
 fixed, precision=4, 1000 sep={\,}, min exponent for 1000 sep=0}
\begin{table}[t] \centering
\pgfplotstabletypeset{
datasetid,vertices,edges,avgoutdegree,maxoutdegree,diameter,size\\
\CR,1.9\,M,2.7\,M,2.8,12,495.0,143\\
\LJ,4.8\,M,68.5\,M,28.3,635\,K,6.5,1617\\
\OR,3.1\,M,117.2\,M,76.3,32\,K,5.0,3066\\
\PA,3.7\,M,16.5\,M,8.7,793,9.4,397\\
\SK,1.7\,M,11.1\,M,13.1,35\,K,5.9,305\\
\TW,40.1\,M,1.4\,B,36.4,2.9\,M,5.4,32686\\
}
\vspace{-5pt}
\caption{Evaluated data sets with their topology statistics.}
\label{tab:data}
\end{table}
}

All evaluated queries are of the form $\{\Set{s},\text{'\emph{*}'},k,k,\rightarrow\}$, where $s$ is a randomly selected start vertex, $*$ refers to a
nonselective edge filter, and $k$ denotes the traversal depth. Without losing generality, we focus in the evaluation on traversal queries where the collection
boundary is equal to the recursion boundary. Such traversal queries only return vertices first discovered in traversal iteration $k$. For the runtime analysis,
we randomly selected start vertices for the traversal and report the median execution time over 50 runs. We decided to report the median since the execution
highly varies for different start vertices.

We compare our \LSFULL and \FIFULL implementations against two join-based approaches (with and without secondary index support) in \SECRET, the open-source
version of Virtuoso Universal Server 7.1~\cite{Erling2012}, and the community edition of the native graph database management system Neo4j 2.1.3~\cite{Neo}. For the
experiments we prepared and configured the evaluated systems as follows:

\paragraph*{SAP HANA} We populated the data sets into two columnar tables, one for vertices and one for the edges. For the indexed join, we created a secondary
index on the source vertex column.

\paragraph*{Virtuoso} Since Virtuoso is an RDF store, we transformed all data sets into RDF triples of the form {\scriptsize \textbf{\texttt{<source\_id>
<edge\_type> <target\_id>}}} and use \abbrev{SPARQL} property paths to emulate a breadth-first traversal. We increased the number of buffers ({\scriptsize
\textbf{\texttt{NumberOfBuffers}}}) and maximum dirty buffers ({\scriptsize \textbf{\texttt{MaxDirtyBuffers}}}) as recommended.

\paragraph*{Neo4j} We configured the object caches of Neo4j so that the data set fits entirely into memory. We warmed up the object cache by running randomly
10000 traversal queries against the database instance. To run the experiments, we used Neo4j's declarative query lanuage Cypher and created an additional index
on the vertex identifier attribute.

\subsection*{TGI Memory Consumption} 
\begin{figure}[t]
\centering
\includegraphics[width=0.48\textwidth,natwidth=610,natheight=640]{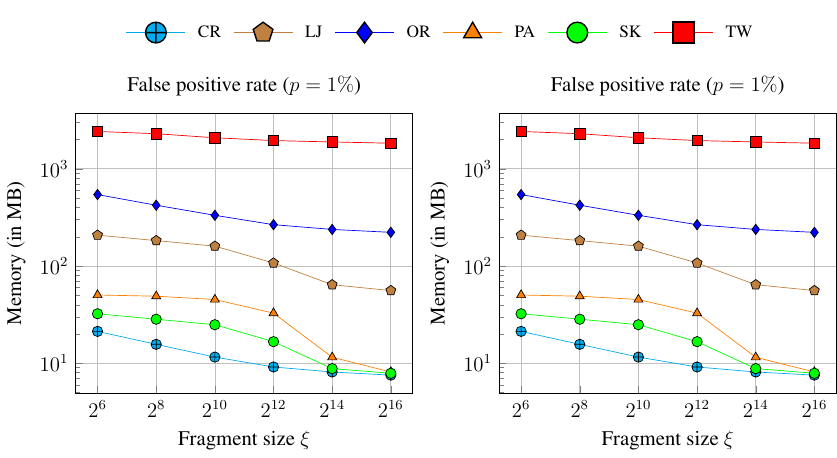}%
\vspace{-5pt}%
\caption{\TGI memory consumption for different false positive rates and fragment sizes.}
\label{exp:tgi_memory_overhead}
\end{figure}

In this experiment we study the effect of various algorithm parameter configurations on the memory consumption of the \TGI.
We populated \TGI instances for clustered physical edge ordering with different fragment sizes and false positive rates, and present the results in
Figure~\ref{exp:tgi_memory_overhead}. To evaluate the impact of the fragment size $\xi$, we construct the \TGI for different fragment sizes $2^6,\ldots,2^{16}$
and a fixed average false positive rate of $1\%$. We analyze the effect of the average false positive rate for a representative fragment size $\xi=512$ and
construct fragment synopses based on an average false positive rate $p$ selected from $\{1\%,5\%,10\%,20\%\}$.

The size of the \TGI is directly related to the total number of edges of the original input graph. For fragment size $\xi=1024$, the \TGI consumes only about
$10\%$ of the size of the input graph. For $\xi=1024$, the \TGI of data set \TW has the highest memory consumption with about
\SI{2.09}{\giga\byte} (about 8.4\% of the raw size of the graph) and data set \CR the lowest memory consumption (about 8.8\% of the raw size of
the graph). For disabled clustering by edge, the memory consumption of the \TGI can grow up to a factor of 10 of the original graph. This makes the unclustered
variant of the \FIFULL impractical for a productive system as it occupies up to two orders of magnitude the memory of the clustered variant.

\paragraph*{Impact of Fragment Size} For all evaluated data sets, the memory footprint decreases for increasing fragment sizes. A larger fragment size leads to
a smaller number of vertices in the \TGI and consequently to fewer possible transitions between them. Although larger fragments cause a denser \TGI topology,
the total number of fragment transitions is much lower than for smaller fragments. For input graphs with a larger average vertex outdegree, the memory overhead
can be reduced for $\xi=2^{16}$ to up to $13\%$ of the memory overhead for $\xi=2^6$. For very sparse graphs, such as \CR, the \TGI consumes
about $37\%$ of the memory for $\xi=2^{16}$ compared to $\xi=2^6$. Consequently, the sparser the input graph is, the lower is the impact of the fragment size on
the total memory consumption of the \TGI.

\paragraph*{Impact of False Positive Rate} We store fragment synopses in space-efficient bloom filter data structures, where each fragment synopsis occupies as
much memory as needed to fulfill the predefined false positive rate. A smaller false positive rate causes the \FIFULL to access more fragments, but reduces the
memory footprint of the \TGI. We show the memory overhead for different false positive rates in Figure~\ref{exp:tgi_memory_overhead}. For data set \PA, a false
positive rate of $20\%$ leads to a memory footprint decrease of $13\%$ compared to a false positive rate of $1\%$. In contrast, data set \CR reached a memory
footprint decrease of almost $50\%$ for $p=20\%$ compared to $p=1\%$.

\begin{figure*}[t!]
\centering
\large
\includegraphics[width=1.0\textwidth,natwidth=610,natheight=640]{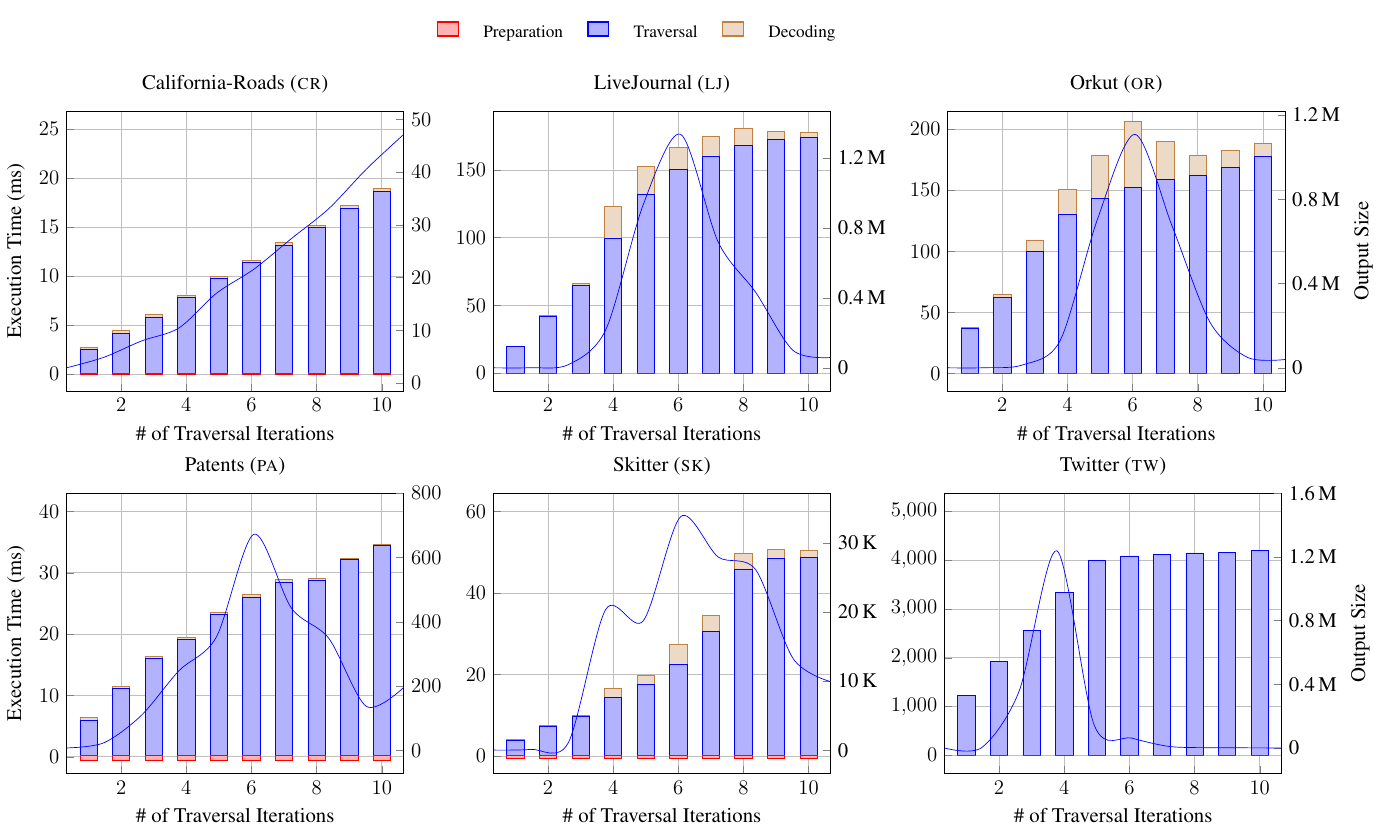}%
\vspace{-5pt}%
\caption{Execution time and output size of \LSFULL for different queries and data sets.}
\label{exp:ls_traversals}
\vspace{-10pt}
\end{figure*}

\subsection*{Runtime Analysis}
Figure~\ref{exp:ls_traversals} presents the runtime results of the \LSFULL for all data sets and with different traversal queries.  We report average execution
times of the three traversal phases \emph{preparation}, \emph{traversal}, and \emph{decoding} as well as average output sizes. In general, the traversal phase
dominates the overall execution time of the traversal operator and consumes up to 95\% of the total runtime. The runtime of the preparation phase is only about
5\% of the overall execution time and is independent from the number of traversal iterations. The preparation phase only evaluates the edge predicate and
processes the start vertices. The decoding phase highly depends on the size of the vertex output set as it translates for each vertex the value code back into
the corresponding vertex identifier. For the data set \SK, we can see the effect of the output size on the runtime spent for the decoding. The output size of
the traversal steadily grows until the traversal reaches the effective diameter. Consequently, only very few traversals reach a larger traversal depth than the
effective diameter. The \LSFULL scales almost linearly with increasing number of traversal iterations as the full column scan takes about the same time to
complete independent of the traversal iteration.

\begin{figure}[t]
\centering
\large
\includegraphics[width=0.5\textwidth,natwidth=610,natheight=640]{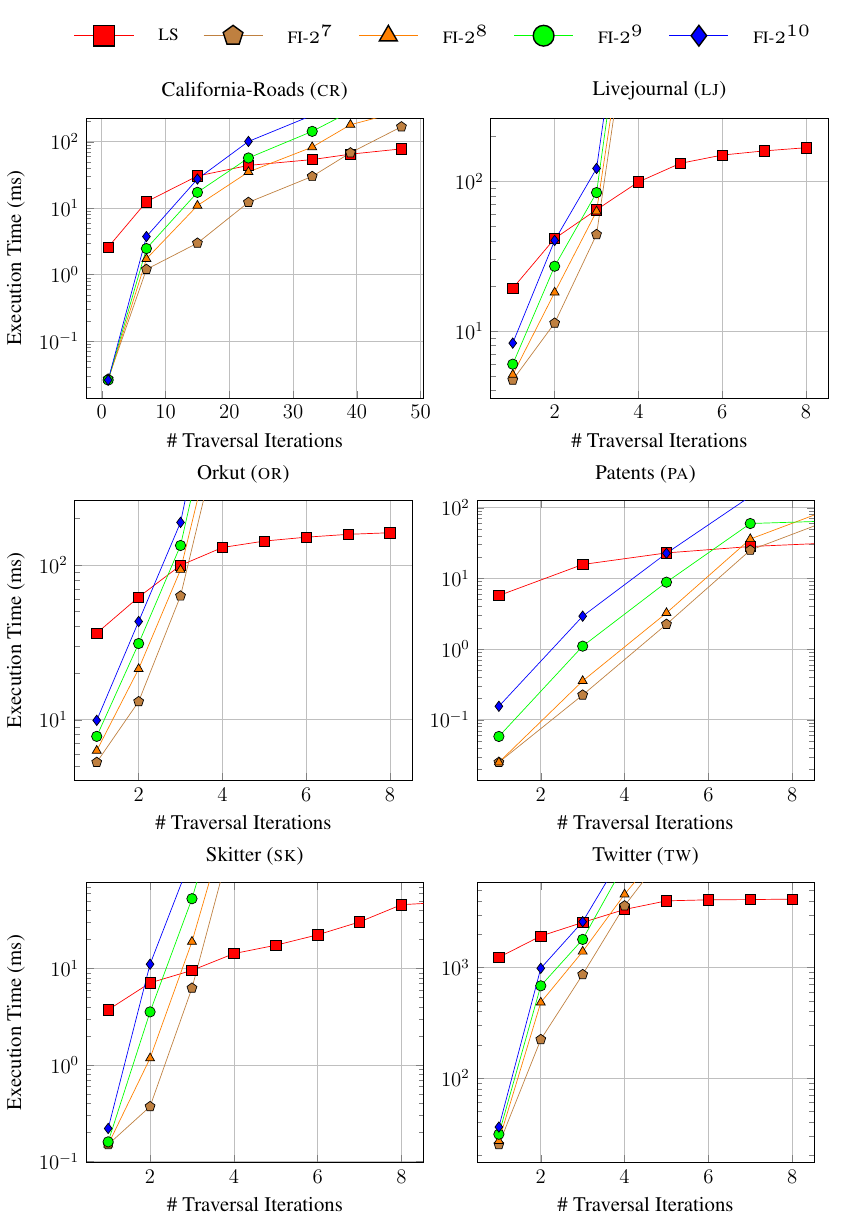}%
\vspace{-5pt}%
\caption{Comparison of \LSFULL and \FIFULL for different queries and data sets.}
\label{exp:comparison_traversals}
\end{figure}

Figure~\ref{exp:comparison_traversals} presents an in-depth comparison of \LSFULL and \FIFULL on all data sets for fragment sizes $\{2^7,\ldots,2^{10}\}$.
Larger fragment sizes resulted in higher execution times and are therefore omitted in the results. For all evaluated data sets, \LSFULL shows a
linear runtime behavior for an increasing number of traversal iterations. For the data set \PA, the runtime steadily increases until the \LSFULL reaches the
effective diameter. After the traversal query reached the effective diameter, the plot flattens for longer traversal queries. In comparison, the data plots of
the \FIFULL grow much faster for an increasing number of traversal iterations. For short traversals with a low number of traversal iterations, the \FIFULL
outperforms the \LSFULL by up to two orders of magnitude. This can be explained by the more fine-granular graph access pattern of the \FI-traversal. Especially
the first traversal iterations process only very small parts of the whole graph and a fine-granular fragment access clearly outperforms a full column scan. For
a large working set, potentially many fragments have to be accessed which in turn is hard to predict and prefetched by the hardware. If large parts of the graph
are accessed, a single full column scan is superior compared to many small fragment scans. The break-even point when the \FIFULL outperforms the \LSFULL depends
on the graph topology and the given traversal query. From the results we can conclude that short traversal queries clearly favor the \FIFULL over the
\LS-traversal. Even for short traversal queries, both data sets produce large intermediate results due to the power-law distribution of vertex outdegrees making
the \FIFULL less effective. For 4 out of 6 data sets, the \FIFULL outperforms the \LSFULL for traversal queries with $r\leq5$. The fragment size has a severe
impact on the overall execution performance of the \FI-traversal. For data set \CR, the fragment size does not only effect the total runtime, but also increases
the range of traversal queries, where the \FIFULL outperforms the \LS-traversal. For example, a traversal query with traversal depth 14 on data set \CR consumes
only about $26\%$ of the runtime than for $\xi=2^{10}$ for a fragment size $\xi=2^7$. In general, we can conclude that \FIFULL is superior to \LSFULL for very
sparse graphs or for short traversal queries.

\begin{figure}[t]
\centering
\includegraphics[width=0.5\textwidth,natwidth=610,natheight=640]{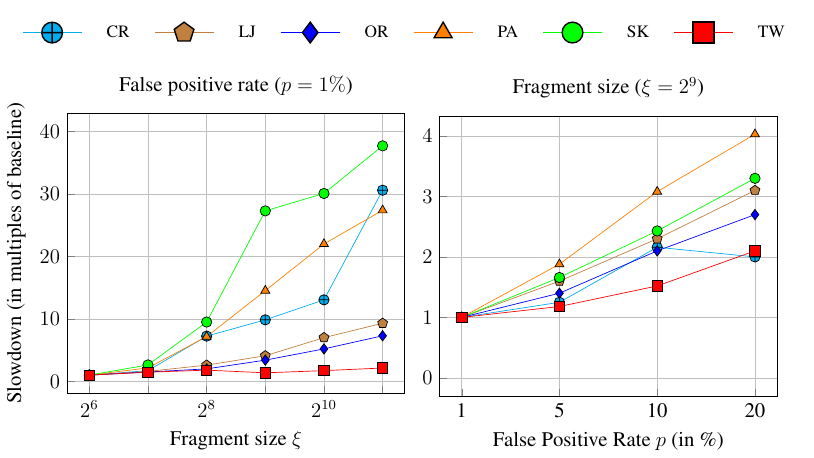}%
\vspace{-5pt}%
\caption{Execution time of \FI-traversal for $\{\Set{s},\text{'\emph{*}'},3,3,\rightarrow\}$.}
\label{exp:tgi_traversal_time}
\end{figure}

Figure~\ref{exp:tgi_traversal_time} depicts the slowdown factors for all data sets with different
fragment sizes $\{2^6,\ldots,2^{16}\}$. To compute the slowdown factor, we use the data point for the smallest fragment size/false positive rate as baseline and
relate all other results to this baseline. Further, we analyze the effect of the false positive rate on the query runtime. Without losing generality, we conduct
all experiments on a representative query of the form $\{\Set{s},\text{'\emph{*}'},3,3,\rightarrow\}$. In general, the \FIFULL with enabled edge clustering
finished the execution on average about 3.5 times faster than for the unclustered variant. If the graph is not clustered by edge, the probability of a
transition to another fragment is significantly higher due to a higher number of distinct values in the fragment. For enabled edge clustering, the maximum
number of possible transitions is bounded by the number of vertices in the graph. For all data sets, smaller fragment sizes close to the expected average vertex
outdegree are more beneficial with respect to execution performance than larger ones. Although one could specify a fragment size that is very small or even
close to 1, the memory overhead would be prohibitively high. Therefore, we limit the minimum fragment size to be connected to the vertex outdegree. \newline
\indent A larger false positive rate increases the memory consumption of the \TGI, but speeds up the runtime of the \FI-traversal. If the false positive rate is
too large, many fragments are read although they do not contribute to the traversal query result.

\begin{figure}[t]
\centering
\large
\includegraphics[width=0.5\textwidth,natwidth=610,natheight=640]{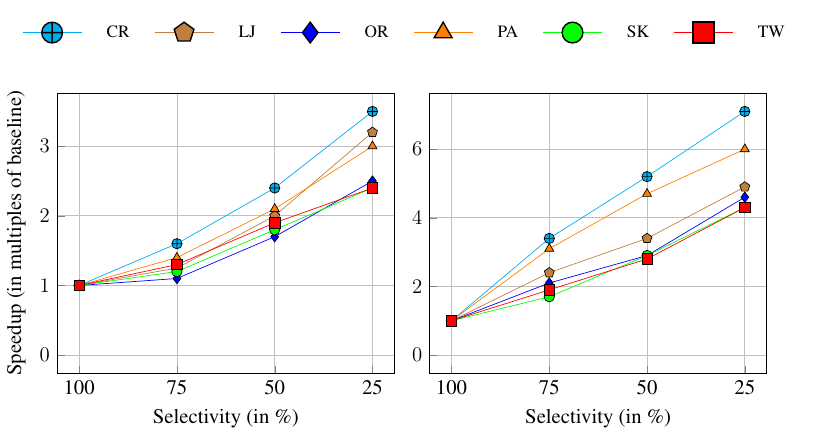}%
\vspace{-5pt}%
\caption{Speedup in multiples of baseline for different edge predicate selectivities with query $\{\Set{s},\text{'\emph{*}'},3,3,\rightarrow\}$. The baseline is a traversal query without edge predicate, i.e., a traversal on the entire graph.}
\label{exp:edge_predicate_traversal_time}
\end{figure}

\paragraph*{Impact of Edge Predicates} We study the effect of edge predicates on the query performance of the \LSFULL and the \FIFULL in
Figure~\ref{exp:edge_predicate_traversal_time}. An edge predicate selects a subgraph of the entire data graph and limits the traversal to a subset of active
edges. We generated edge weights following a zipfian distribution with $s=2$ and assigned them randomly to the edges. For a selectivity of $25\%$, i.e., an edge
predicate that selects only $25\%$ of all edges leads for the \LSFULL to a speedup of 3. We observed that an edge predicate with a low selectivity drastically
reduces the size of intermediate and final output results. Since the \LSFULL is a scan-based traversal algorithm, it still has to scan the entire column for
each traversal iteration. In contrast, the \FIFULL reaches a speedup of up to factor 6 for a selectivity of $25\%$. If the selectivity is low, more fragments
can be pruned during the traversal and cause the doubled speedup compared to the \LSFULL.

\begin{figure}[t!]
\centering
\large
\includegraphics[width=0.5\textwidth,natwidth=610,natheight=640]{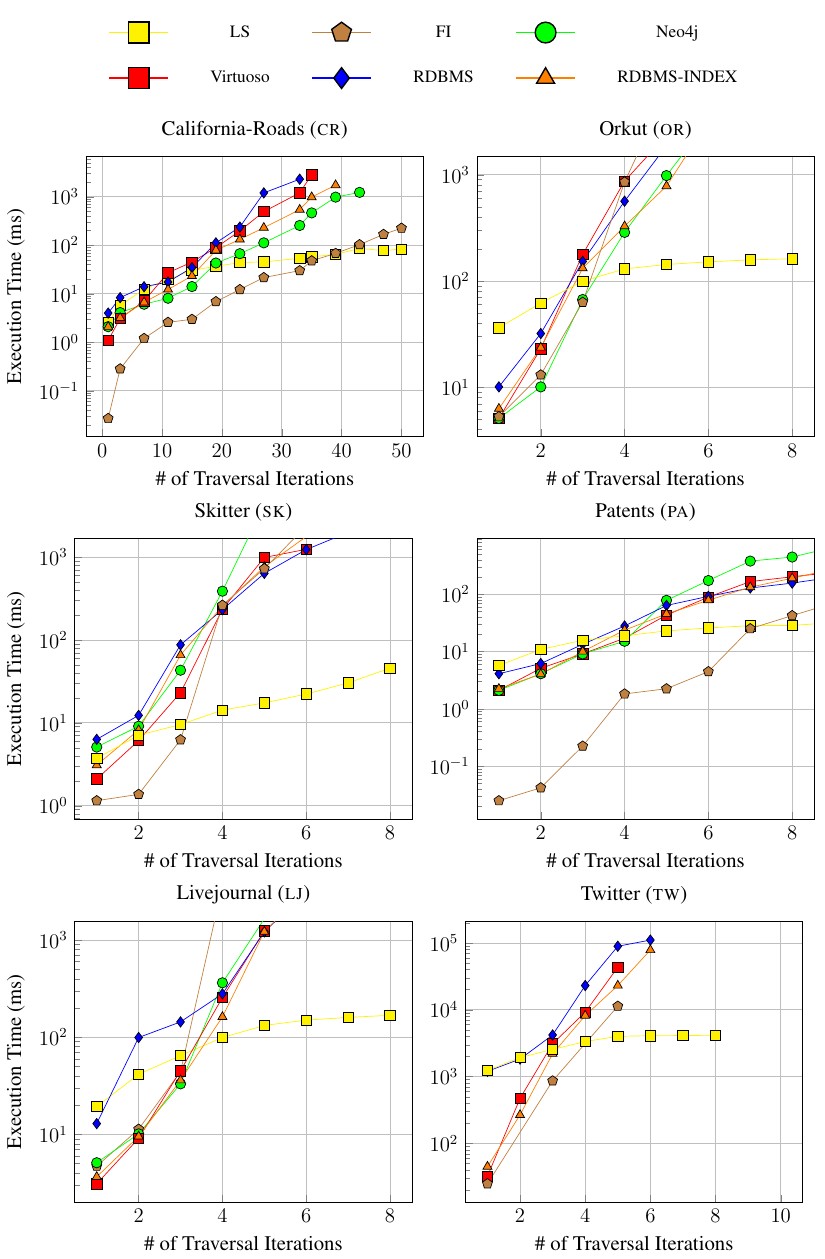}
\vspace{-5pt}
\caption{System-level benchmarks of \LS-traversal, \FI-traversal, \abbrev{Virtuoso}, \abbrev{Neo4j}, and a self-join-based approach (with and without secondary index support).
Since the execution times highly vary for different start vertices, we report
median execution times. We were not able to run experiments for the Twitter data set on Neo4j due to data loading issues.}
\label{exp:system_comparison_traversals}
\end{figure}

\paragraph*{System-Level Benchmarks} We compared our two traversal implementations with a purely relational self-join-based approach (with and without secondary
index support), Neo4j, and Virtuoso 7.0. For the join-based traversal, we use the same data layout as for \GRAPHITE and leverage the columnar relational engine
of \SECRET. We present our results in Figure~\ref{exp:system_comparison_traversals}. For short traversals of 1--3 hops, our \FIFULL is competitive with native
graph implementations from Neo4j and Virtuoso. For data sets \PA, \SK, and \CR \FIFULL outperforms all evaluated systems by up to an order of magnitude.

\subsection*{Cost Model Evaluation} To verify our cost model function, we applied regression analysis. We use the \emph{coefficient of determination} denoted
$R^2$ to evaluate the quality of our \FIFULL cost model. The coefficient of determination has a value range of $ -1 \leq |R^2| \leq 1$. A value of $|R^2|$ close
to 1 indicates a good fit of the proposed cost model function with the manually collected data points. We compare the manually collected data of the number of
accessed edges against the results of the cost function. For each data set, we performed a set of traversal queries with a recursion boundary ranging from 1 to
10. We achieved the best result with an average $R^2=0.92$ for data set \CR, respectively. For the data set \EP, we achieved $R^2=0.78$. In general, graphs with
a power-law vertex outdegree distribution caused our cost function to underestimate the costs of the \FI-traversal. This underestimation can be explained by the
method used to describe the vertex outdegree distribution. We use the average vertex outdegree $\bar{d}_{\text{out}}$ to estimate the expected number of
neighbors for a single vertex. If the traversal discovers a vertex with a considerably larger vertex outdegree, the cost function underestimates the access
costs. Additionally, the traversal depth is estimated as the minimum of the recursion boundary and the diameter of the graph. However, traversal queries that
terminate before they reach the recursion boundary, are not appropriately reflected in the cost function. We believe that additional information about the
outdegree distribution and the distribution of path lengths is required to obtain a more accurate estimation from the cost function.

%% file: 8-related.tex
\section{Related Work}
\label{sec:related}

\noindent \paragraph*{Graph Traversal Algorithms} Graph traversals are one of the most important and fundamental building blocks of graph algorithms, such as
finding shortest paths, computing the maximum flow, and identifying strongly connected components. Increasing graph data sizes and the proliferation of
parallelism on different hardware levels as well as heterogeneous processor environments encouraged researchers to revise the well-known breadth-first graph
traversal and to propose novel techniques to run graph traversals on high-end computers with large numbers of cores and different types of processors on a
single machine. A large body of research has been conducted on efficient parallel graph traversals, lately even leveraging co-processors to speed up graph
processing on large data graphs~\cite{Hong2011}. State-of-the-art parallel graph traversals operate with a level-synchronous strategy and parallelize the work
to be done at each level. However, all parallel graph traversal implementations rely on sophisticated data structures that are tailored to the graph traversal
algorithm. Such an algorithm-dependent data structure is not applicable in our case since we are using the traversal operator in an \RDBMS on top of a common
storage engine without copying data into separate data structures. As one of our strongest advantages, we do not require the graph data to be copied from
possibly already existing legacy relational tables into algorithm-specific data structures. Replicating data into separate data structures wastes memory and
also adds a considerable maintenance overhead. Chhugani et al. study scalable breadth-first traversal algorithms on modern hardware with multi-socket,
multi-core processor architectures~\cite{Chhugani2012,Prabhakaran2012}. They achieved an impressive performance by tuning the data structures and the traversal
algorithm to the underlying hardware. In contrast to our approach, they only consider a single implementation of graph traversals for any graph topology and
types of graph traversal queries. Graph traversals in distributed memory recently gained more attention and resulted in the development of sophisticated data
partitioning schemes for distributed graph traversals~\cite{Buluc2011,Pearce2010}.

\paragraph*{Graph Processing on Column Stores} Column stores have shown great potential for storing and querying wide and sparse data~\cite{Abadi2007a}. These
considerations brought up research projects that aimed to provide efficient access to \RDF~\cite{Abadi2007} and \XML data~\cite{Teubner2006} kept in a column
store. However, none of them covered the design and implementation of a native graph traversal operator that leverages advantages of columnar data structures
and exploits knowledge about the graph topology to speed up the graph traversal execution.

\paragraph*{Distributed Graph Engines} The demand to efficiently process real-world billion-scale graphs triggered the development of a variety of distributed
graph processing systems~\cite{Gonzalez2012,Kang2011,Low2012,Malewicz2010}. \GBASE is a distributed graph engine based on MapReduce and relies on distributed
matrix-vector multiplications~\cite{Kang2011}. The vertex-centric programming model, as proposed by \cite{Malewicz2010}, has been an area of active research and
has been implemented in GraphLab~\cite{Low2012} and PowerGraph~\cite{Gonzalez2012} among others. Although distributed graph engines show good scalability for
billion-scale graphs, we see the following disadvantages making them not applicable in our scenarios: (1) Business data from enterprise-critical applications is
still mainly stored in \RDBMS and cannot be easily replicated to external graph processing engines; (2) typically, graph engines cannot cope with
cross-data-model operations (e.g., combining text, graph, and spatial); (3) distributed graph engines rely on sophisticated graph partitioning algorithms that
do not scale well to large graphs and are hard to maintain for dynamic graphs; and (4) they do not provide transactional guarantees.

\paragraph*{Single-Machine Graph Engines} An interesting alternative to distributed graph engines has been introduced by Kyrola et al.~\cite{Kyrola2012} and
conceptually extended by Han et al.~\cite{Han2013}. \GRAPHCHI is a disk-based graph engine on a single machine that exploits parallel sliding windows and
sharding to efficiently process billion-scale graphs from disk~\cite{Kyrola2012}. To minimize I/O overhead, they apply a technique similar to edge clustering to
improve disk access and maximize data locality on disk. The lack of support for attributes on vertices and edges and dynamic graphs resulted in GraphChi-DB, a
recent extension of GraphChi~\cite{Kyrola2014}. Interestingly, they also use a vertically partitioned layout to represent attributes on vertices and edges. In
contrast to \GRAPHCHI we run \GRAPHITE not as a standalone graph engine, but as part of a graph runtime stack on a common relational storage engine in an
\RDBMS. Since our targeted scenarios run on main-memory \RDBMS, \GRAPHITE can operate completely in memory and aims at maximizing \CPU cache locality. Similar
to \GRAPHCHI is \TURBOGRAPH, a single-machine disk-based graph processing engine using solid state disks~(\SSD) to store and process large graphs~\cite{Han2013}.
They use a vertically partitioned layout on \SSD to store vertex attributes. However, we do see two major drawbacks compared to our approach: (1) \GRAPHCHI and
\TURBOGRAPH\ are efficient single-machine graph engines, but do not provide transactional access to the data; and (2) graph data has to be available upfront in
a specific data format on disk. To be applicable for business data stored in \RDBMS, the data has to be exported and transformed into a file format that can be
consumed by the system.

\paragraph*{Graph Databases} A different direction is followed by graph data\-bases, such as Neo4j~\cite{Neo}, Sparksee~\cite{Martinez-Bazan2012}, and
InfiniteGraph~\cite{Inf}. While Neo4j relies on a disk-based storage accelerated by buffer pools to store recently accessed parts of the graph, Sparksee allows
manipulating and querying the graph in memory. The Sparksee internal data structures rely on efficient bitmaps, which represent the set of vertices and edges
describing the graph~\cite{Martinez-Bazan2012}. All graph databases are specialized engines that can perform graph-oriented processing efficiently, but always
require loading possibly relational business data in advance from other data sources. On the contrary, our approach can directly operate on the relational
business data without having to copy it to a dedicated database engine. Moreover, the combination of relational operations and graph operations can be handled
efficiently by a single database engine.

\paragraph*{Graph Processing in \RDBMS} Although graph databases are a rather new research field, path traversals in relational databases with the help of
recursive queries have been in the focus of research for more than 20 years now~\cite{Agrawal1988}. There have been proposals for extending relational query
languages with support for recursion in the past and even the SQL:1999 standard offers recursive common table expressions. However, commercial database vendors
often provide their own proprietary functionality, if they do at all. Gao et al. leverage recent extensions to the SQL standard, such as \emph{window functions}
and \emph{merge statements}, to implement algorithms for shortest path discovery on relational tables~\cite{Gao2011}. Unlike our approach, they are reusing
existing relational operators and the relational query optimizer to create an optimal execution plan. However, when ignoring graph-specific statistics, the
optimizer is likely to select a suboptimal execution plan. Magic-set transformations are a query rewrite technique for optimizing recursive and non-recursive
queries, which was originally devised for Datalog~\cite{Bancilhon1986} and has been extended for SQL~\cite{Mumick1994}. Since graph traversals can be expressed
as recursive database queries, the magic-sets transformation could also be applied to them. However, instead of proposing an optimization strategy for
relational execution plans, we approach the problem with a dedicated plan operator.

%% file: 9-conclusion.tex
\section{Conclusion}
\label{sec:summary}

\noindent We presented \GRAPHITE, a modular and versatile graph traversal framework for main-memory \RDBMS. As part of \GRAPHITE, we presented two different
specialized traversal implementations named \LSFULL and \FIFULL to support a wide range of different graph topologies and varying graph traversal queries most
efficiently. \GRAPHITE is extensible and other graph traversal strategies, such as depth-first based traversals, could be integrated as well.
We derived a basic cost model of two traversal implementations and experimentally showed that it can assist a query optimizer to select the optimal traversal
implementation. The \FIFULL outperforms the \LSFULL for graphs with a low density and short traversal queries by up to two orders of magnitude. In contrast, the
\LSFULL performs significantly better than the \FI-traversal, if the graph is dense or the query traverses a large fraction of the whole graph.
Our experimental results illustrate the need for a graph traversal framework with an accompanied set of traversal operator implementations. Finally, we show
that, despite popular belief, graph traversals can be efficiently implemented in \RDBMS on a common relational storage engine and are competitive with those of
native graph management systems.